\documentclass[prd,preprint,tightenlines,floatfix,preprintnumbers,nofootinbib]{revtex4}
 \usepackage[dvips,final]{graphicx}
  \usepackage{amssymb,amsmath,bm,pifont}

\begin{document}
\hfill\hbox{RUB-TPII-02/09}
\title{Pion Form Factor in the NLC QCD SR approach
\footnote{
          Talk presented by the second author
          at the Workshop on Physics of Fundamental Interactions,
          Institute of High Energy Physics,
          Protvino, Russia, 22--25 December 2008.
          }
}

\author{\firstname{A.~P.}~\surname{Bakulev}}
\email{bakulev@theor.jinr.ru}

\author{\firstname{A.~V.}~\surname{Pimikov}}
\email{pimikov@theor.jinr.ru}
\affiliation{%
Bogoliubov Laboratory of Theoretical Physics,
JINR, 141980 Dubna, Russia
            }%

\author{\firstname{N.~G.}~\surname{Stefanis}}
\email{stefanis@tp2.ruhr-uni-bochum.de}
\affiliation{%
Institut f\"{u}r Theoretische Physik II,
Ruhr-Universit\"{a}t Bochum, D-44780 Bochum, Germany
            }%

\begin{abstract}
We present results of a calculation of the electromagnetic pion
form factor within a framework of QCD Sum Rules with nonlocal
condensates and using a perturbative spectral density which includes
$\mathcal{O}(\alpha_s)$ contributions.
\end{abstract}
\maketitle

\section{Introduction}
An archetypical example of a QCD (hadronic) observable is the pion
form factor, which is typical for a hard-scattering process obeying
a factorization theorem \cite{ER80tmf,LB80}.
Consequently,
at asymptotically large $Q^2$
it can be cast in terms of a scale-dependent pion
distribution amplitude (DA)~\cite{Rad77} of leading twist two
$\varphi_{\pi}(x,Q^2)$ convoluted with the hard-scattering amplitude of
the process which contains the large external scale $Q^2$:
\begin{eqnarray}
 \label{eq:FF.pi.Fact}
  F^{\text{pert}}_{\pi}(Q^2)
   = \frac{8\pi \alpha_s(Q^2) f_{\pi}^2}{9\,Q^2}
      \left|I^\pi_{-1}(Q^2)\right|^2
  ~~~\text{with}~~~
  I^\pi_{-1}(Q^2)
   = \int\limits_0^1\frac{\varphi_{\pi}(x,Q^2)}{x}\,dx \,.
\end{eqnarray}
The nonperturbative input---the pion DA $\varphi_{\pi}(x,\mu^2)$---can
be expressed as an expansion over Gegenbauer polynomials
\begin{eqnarray}
 \varphi_{\pi}(x,\mu^2)
  = \varphi^\text{as}(x)
     \left[ 1
          + \sum\limits_{n\geq1}a_{2n}(\mu^2)\, C_{2n}^{3/2}(2x-1)
     \right],~~~
 I^\pi_{-1}(\mu^2)
  = 3\left[1 + \sum\limits_{n\geq1}a_{2n}(\mu^2)
       \right],~~~
 \label{eq:Gegen}
\end{eqnarray}
where the asymptotic pion DA has the form
\begin{eqnarray}
 \varphi^\text{as}(x)
  = 6\,x\,(1-x)\,,
\end{eqnarray}
while the scale dependence
of coefficients $a_{2n}(\mu^2)$
is controlled by the ERBL evolution
equation \cite{ER80tmf,LB80}.

At the one-loop level and at asymptotically large $Q^2$,
the pion form factor simplifies to
$
F^\text{pert}_{\pi}(Q^2) = 8\,\pi\,\alpha_s(Q^2)\,f_{\pi}^{2}/Q^2\
$
The onset of the asymptotic regime cannot be determined precisely;
estimates~\cite{BPSS04,BLM07} show that this transition scale is
of the order of $100$~GeV$^2$.

On the other hand, at intermediate momentum transfers
$20~{\rm GeV}^2 \geq Q^2\geq1$~GeV$^2$, the situation is more
complicated because of the interplay of perturbative and
nonperturbative effects.
The latter effects are contained in a non-factorizable part---called
the soft contribution---so that one has to take it into account using
some nonperturbative concepts, e.g.,
the method of QCD sum rules (SR) \cite{NR82,IS82,BR91},
the local quark-hadron duality (LD) approach~\cite{NR82,Rad95},
and others.
Note in this context that,
describing the pion form factor within
the three-point QCD SR approach \cite{NR82,IS82},
the shape of the pion DA becomes irrelevant.
This considerably reduces the inherent theoretical uncertainty of the
method.
The same applies to the LD approach, but the latter contains an
additional uncertainty related to the $s_0(Q^2)$ setting for
intermediate and large values of $Q^2$---see for a discussion in
\cite{BLM07}.

However, the standard QCD SR~\cite{NR82,IS82} are plagued by
instabilities arising at $Q^2\gtrsim3$~GeV$^2$, which are induced by
those terms in the operator product expansion that are either constant
or grow linearly with $Q^2$ (see Tab.\ \ref{tab:History}).
Such terms do not represent a nonperturbative contribution correctly.
The corresponding diagrams result from the substitution of some
propagators by constant factors that denote condensates lacking a
$Q^2$-dependence, viz.,
$\langle T(q(z)\bar q(0))\rangle
\rightarrow\langle \bar q(0)q(0)\rangle$.
The scale dependence is retrieved by including in the calculation the
contributions stemming from higher-dimension operators, like
$\langle \bar q(0)D^2q(0)\rangle$,
$\langle \bar q(0)(D^2)^2q(0)\rangle$
\textit{etc.},
that are entailed by the Taylor expansion
of the original nonlocal condensate (NLC),
$\langle \bar q(0)q(z)\rangle$,
being the nonperturbative part of the quark propagator.
In order to obtain the correct large-$Q^2$ behavior and ensure
that the total condensate contribution decreases for large $Q^2$,
one has to resum all terms of the standard OPE
bearing terms of the sort
$(Q^2/M^2)^n$.
This is a rather tedious task and, therefore, we refrain from using
the original Taylor expansion in our analysis and take instead
recourse to a modified diagrammatic technique which makes use of new
lines and vertices associated with NLC (details can be found
in~\cite{BR91}).
\begin{table}[bh]\vspace*{-3mm}
\caption{$Q^2$-behavior of the nonperturbative contribution in different QCD SR
 approaches.
 Here, $c_1,~c_2,~c_3,~c_4$ are dimensionless constants (not depending on
 $Q^2$).
 The abbreviations used are: LD for local duality, LO for leading order,
 and NLO for next-to-leading order, while $\lambda_q^2$ and $M^2$ denote
 the vacuum quark virtuality and the Borel parameter, respectively.
   \label{tab:History}\vspace*{+1mm}}
\begin{ruledtabular}
 \begin{tabular}{l|c|c|l}
 {\strut\vphantom{\vbox to 6mm{}}} $_{\vphantom{\vbox to 4mm{}}}$
 Approach                               & Accuracy~ & Condensates
& $Q^2$-behavior of $\Phi_{\text{OPE}}$                              \\ \hline
{\strut\vphantom{\vbox to 6mm{}}} $_{\vphantom{\vbox to 4mm{}}}$
 Standard QCD SR~\cite{NR82,IS82}       & LO       & Local
& $c_1 + Q^2/M^2$                                                    \\ \hline
{\strut\vphantom{\vbox to 6mm{}}} $_{\vphantom{\vbox to 4mm{}}}$
 QCD SR with NLCs~\cite{BR91}           & LO       & Local $+$ Nonlocal~
& $\left(c_2+Q^2/M^2\right)\left(e^{- c_3 Q^2\lambda_q^2/M^4}+c_4\right)$
                                                                     \\ \hline
{\strut\vphantom{\vbox to 6mm{}}} $_{\vphantom{\vbox to 4mm{}}}$
 LD SR($M^2\to \infty$)~\cite{BLM07}~ & NLO      & ---
& $0$                                                                \\ \hline
{\strut\vphantom{\vbox to 6mm{}}} $_{\vphantom{\vbox to 4mm{}}}$
 This paper                             & NLO       & Nonlocal
& $\left(c_1 + Q^2/M^2\right)\,e^{- c_3 Q^2\lambda_q^2/M^4}$
 \end{tabular}
\end{ruledtabular}
\end{table}

An earlier attempt to generalize the QCD SR~\cite{BR91} approach by
employing such NLC contributions turned out to be incomplete, because
it was found to contain contributions originating from local
condensates.
This is related to the fact that a specific model
(\ref{eq:fi.barqAq.BR91}) for the 3-point quark-gluon-antiquark NLC was
used in which the NLC $M_i(x^2,y^2,(x-y)^2)$ are nonlocal only with
respect to one single separation, say, $x^2$, out of the three possible
inter-parton separations $x^2$, $y^2$, and $(x-y)^2$.
As a result, also this type of approach suffers from the same
shortcomings as the standard QCD SR.
In contrast,
LD SR have no condensate contribution
due to the $M^2\rightarrow \infty$ limit.
The only trace of all contributing condensate contributions is
embodied in the parameter $s_0$, which can be derived from
the LD sum rule for $f_{\pi}$.
Due to the Ward identity, these sum rules are connected only at
$Q^2=0$, so that the applicability of this method to the pion form
factor is actually confined to low momenta around $Q^2\ll s_0$.
The definition of $s_0$ at large $Q^2$ is not settled
in this approach~\cite{BLM07}.

In this presentation, we report upon an investigation of the
electromagnetic pion form factor which employs QCD SR with NLC
\cite{BR91,Rad95}.
This enables us to enlarge the region of applicability of the QCD SR to
momenta as high as $10~\text{GeV}^2$.
Moreover, we use a spectral density which includes terms of
$\mathcal{O}(\alpha_s)$.
The influence of this NLO contribution to the pion form factor reaches
the level of $20\%$.
The remainder of this report is organized as follows.
The next section contains the necessary ingredients of the QCD SR
approach with NLC.
The second part of this section contains also our results.
Our conclusions are given in section \ref{sec:Conclusions}, where we further
discuss our findings in comparison with the available experimental data,
lattice simulations, and other theoretical calculations.

\section{Pion form factor from QCD sum rules with nonlocal condensates}
 \label{sec:NLC.QCD.SR.FF}
The nonlocality of the QCD vacuum, suggested in
\cite{MR86,MS93,BM98,BMS01}, is crucial for a correct determination of
the pion DA and the computation of the pion form factor
\cite{BR91,Rad95}.
For that reason, let us first recall the main elements of this
approach and discuss its application to the calculation of 3-point
correlators in QCD.

For the scalar and vector condensates, we employ the same Gaussian model
as in~\cite{BM98,BMS01}, i.e.,
\begin{eqnarray}
\label{eq:S.V.NLC}
 \langle{\bar{q}(0)q(z)}\rangle
=
 \langle{\bar{q}q}\rangle\,
     e^{-|z^2|\lambda_q^2/8}\,;\;
 \langle{\bar{q}(0)\gamma_\mu q(z)}\rangle
=
 \frac{i\, z_\mu\,z^2}{4}\,
    A_0\
      e^{-|z^2|\lambda_q^2/8}\,,
\end{eqnarray}
where $A_0=2\alpha_s\pi\langle{\bar qq}\rangle^2/81$.
Note that above, a Fock--Schwinger string is attached in-between the
quark-antiquark fields in order to preserve gauge invariance.
But adopting the fixed-point (Fock--Schwinger) gauge
$z^\mu A_\mu(z)=0$
each string reduces to unity, provided the integration path in
the exponent is a straight line going from $0$ to $z$.
The nonlocality parameter $\lambda_q^2 = \langle{k^2}\rangle$
provides a useful measure of the average momentum of quarks in the
QCD vacuum.
It has been estimated in QCD SR~\cite{BI82,OPiv88} and on the
lattice~\cite{DDM99,BM02} with a value around
$\lambda_q^2 = 0.45\pm 0.1\text{~GeV}^2$.
To parameterize the vector (V) and the axial-vector (A)
quark-gluon-antiquark condensate,
we use the expressions derived in \cite{MR86}:
\begin{eqnarray}
\langle{\bar{q}(0)\gamma_\mu(-g\widehat{A}_\nu(y))q(x)}\rangle
&=&
   (y_\mu x_\nu-g_{\mu\nu}(y\cdot x))\overline{M}_1(x^2,y^2,(y-x)^2)
\nonumber\\
&+&
   (y_\mu y_\nu-g_{\mu\nu}y^2)\overline{M}_2(x^2,y^2,(y-x)^2)\,,
\nonumber
\\
\langle{\bar{q}(0)\gamma_5\gamma_\mu(-g\widehat{A}_\nu(y))q(x)}\rangle
&=&
   i\varepsilon_{\mu\nu yx}\overline{M}_3(x^2,y^2,(y-x)^2)
\vspace{-5mm}\nonumber
\end{eqnarray}
with
\begin{eqnarray}
 \overline{M}_i(x^2,y^2,z^2)
  = A_i\int\!\!\!\!\int\limits_{\!0}^{\,\infty}\!\!\!\!\int\!\!
        d\alpha \, d\beta \, d\gamma \,
         f_i(\alpha ,\beta ,\gamma )\,
          e^{\left(\alpha x^2+\beta y^2+\gamma z^2\right)/4}\,,
\label{eq:M}
\end{eqnarray}
where the following abbreviation
$A_{1,2,3}\equiv A_0 \times\left(-\frac32,2,\frac32\right)$
was used.
The minimal Gaussian model of the nonlocal QCD vacuum is introduced
by the following ansatz
\begin{eqnarray} \label{eq:Min.Anz.qGq}
f^\text{min}_i\left(\alpha,\beta,\gamma\right)
   = \delta\left(\alpha -\Lambda\right)\,
     \delta\left(\beta  -\Lambda\right)\,
     \delta\left(\gamma -\Lambda\right)
\end{eqnarray}
with $\Lambda=\lambda_q^2/2$.
This model violates the QCD equations of motion, while at the same time
the 2-point correlator of the vector currents is not transverse.
To restore the QCD equations of motion and to minimize the
non-transversity of the $VV$ correlator, an improved model
of the QCD vacuum was proposed \cite{BP06}:
\begin{eqnarray}\label{eq:Imp.Anz.qGq}
f^\text{imp}_i\left(\alpha,\beta,\gamma\right)
  = \left(1 + X_{i}\partial_{x}
            + Y_{i}\partial_{y}
            + Y_{i}\partial_{z}
    \right)
         \delta\left(\alpha-x\Lambda\right)
         \delta\left(\beta -y\Lambda\right)
         \delta\left(\gamma-z\Lambda\right)\,.
\end{eqnarray}
Here $\Lambda=\frac12\lambda_q^2$ and $z=y$, whereas
\begin{subequations}
\begin{eqnarray}
  X_1 &=& +0.082\,;~X_2 = -1.298\,;~X_3 = +1.775\,;~x=0.788\, ,~~~\\
  Y_1 &=& -2.243\,;~Y_2 = -0.239\,;~Y_3 = -3.166\,;~y=0.212\,.~~~
\end{eqnarray}
\end{subequations}
These parameters satisfy the supplementary conditions
\begin{eqnarray}
 \label{eq:3L.D.A.Rule4}
  12\,\left(X_{2} + Y_{2}\right)
  - 9\,\left(X_{1} + Y_{1}\right)
  = 1 \,,~~~~~x+y=1\, ,
\end{eqnarray}
following from the QCD equations of motion.

The Borel SR for the pion form factor, based on the 3-point
AAV correlator, was considered for local condensates in
\cite{NR82,IS82}, whereas the NLC case was treated in \cite{BR91}.
This way, one obtains the following SR
\begin{eqnarray}
\label{eq:ffQCDSR}
  f_{\pi}^2\,F_{\pi}(Q^2)
=
  \int\limits_{0}^{s_0}\!\!\int\limits_{0}^{s_0}\!ds_1\,ds_2\
           \rho_3(s_1, s_2, Q^2)\,
           e^{-(s_1+s_2)/M^2}
     + \Phi_{G}(Q^2,M^2)
     + \Phi_{\langle\bar{q}q\rangle}(Q^2,M^2)\,.
\end{eqnarray}
Note that as long as the condensate terms $\Phi_{G}$ and
$\Phi_{\langle\bar{q}q\rangle}$ are not specified, this SR
may have a local or nonlocal content.
On the other hand, the perturbative 3-point spectral density is given
by
\begin{eqnarray}
 \label{eq:SpDen.pert}
  \rho^{(1)}_3(s_1, s_2, Q^2)
  &=& \left[\rho_3^{(0)}(s_1, s_2, Q^2)
        + \frac{\alpha_s(Q^2)}{4\pi}\,
           \Delta\rho_3^{(1)}(s_1, s_2, Q^2)
    \right]\,.
\end{eqnarray}
The leading-order spectral density $\rho_3^{(0)}(s_1, s_2, t)$
is known since the early eighties~\cite{NR82,IS82}.
As regards the next-to-leading order (NLO) spectral density
$\Delta\rho_3^{(1)} (s_1, s_2, Q^2)$, it has been obtained quite
recently \cite{BO04}.
The phenomenological side of the SR contains the contribution
which stems from higher resonances, modeled via the
spectral density
\begin{equation}
 \label{eq:HR}
  \rho_\text{HR}(s_1, s_2)
  =  \left[1-\theta(s_1<s_0)\theta(s_2<s_0)\right]\,
      \rho_3(s_1, s_2, Q^2)
\end{equation}
and using the continuum-threshold parameter $s_0$.
In order to improve the low-scale behavior of the pion form factor,
we adopt a scheme, developed in \cite{BRS00,BPSS04}, and employ
an analytic running coupling \cite{SS97}
\begin{eqnarray}
 \label{eq:alphaS}
  \alpha_s(Q^2)
   &=&\frac{4\pi}{b_0}
       \left(\frac{1}{\ln(Q^2/\Lambda_{\text{QCD}}^2)}
           - \frac{\Lambda_{\text{QCD}}^2}{Q^2-\Lambda_{\text{QCD}}^2}
                \right)\,,
\end{eqnarray}
with $b_0=9$ and $\Lambda_\text{QCD}=300$~MeV.

For our discussion to follow, we use for the nonperturbative terms
$\Phi_{G}$ and $\Phi_{\langle\bar{q}q\rangle}$ in the
local-condensate case the following expressions \cite{NR82,IS82}:
\begin{eqnarray}
 \label{eq:Phi.G.qq.Loc}
 \Phi_{G}^\text{loc}(M^2)
  = \frac{\langle\alpha_s GG\rangle}{12\,\pi\,M^2}
 \,,~~~~~~
 \Phi_{\langle\bar{q}q\rangle}^\text{loc}(Q^2,M^2)
  = \frac{104\,A_0}{M^4}\,
     \left(1+\frac{2\,Q^2}{13\,M^2}\right)\,.
\end{eqnarray}
These expressions, that are used in the standard QCD SR~ for
the pion form factor, show a wrong behavior at large $Q^2$:
(i) The quark contribution contains both a linearly growing
term as well as a constant one. (ii) The gluon contribution
is just a constant.
On the other hand, the perturbative term on the right-hand side
of Eq.\ (\ref{eq:ffQCDSR}) behaves at large $Q^2$
like $s_0/Q^4$ or $M^2/Q^4$.
Hence, the SR becomes unstable for $Q^2>3$~GeV$^2$.
But using the generalized QCD SR with NLC, \cite{BR91},
this deficiency should not appear.
Alas, even this approach has a dark side,
because it still uses in the analysis
of the pion form factor
a partially local parameterization of the quark-gluon-antiquark NLC.
Indeed, the following parametric functions (\ref{eq:M})
have been used in~\cite{BR91}
($\Lambda=\lambda_q^2/2$):
\begin{eqnarray}
\label{eq:fi.barqAq.BR91}
 f^\text{BR}_i\left(\alpha,\beta,\gamma\right)
&=&
         \delta\left(\alpha-x_{i1}\Lambda\right)
         \delta\left(\beta -x_{i2}\Lambda\right)
         \delta\left(\gamma-x_{i3}\Lambda\right)\,,
\\\nonumber
 x_{ij}
&=&
   \left(
        \begin{array}{ccc}
           0.4 & 0   & 0.4\\
           0   & 1   & 0.4\\
           0   & 0.4 & 0.4
        \end{array}
  \right)\,.
\end{eqnarray}
The absence of nonlocality effects in (\ref{eq:M})
for the quark-antiquark separation $y^2$ ($i=1$)
and also
for the (anti)quark-gluon separations $x^2$ and $(x-y)^2$
($i=2,3$) is revealed by the zero elements
in the matrix $x_{ij}$.

Note that the NLC contributions to the pion form factor, entering
the SR (\ref{eq:ffQCDSR}), can still be used in connection with
an improved version of the quark-gluon NLC
because the expressions
obtained in \cite{BR91} have the form of a convolution
in the $\alpha$-representation of the NLC distribution functions
$f_i(\alpha,\beta,\gamma)$ with model-independent coefficient
functions.
In the present work we apply
the minimal (\ref{eq:Min.Anz.qGq})
and the improved (\ref{eq:Imp.Anz.qGq}) Gaussian models of NLC.
The contribution from the vector condensate
to $\Phi_{\langle\bar{q}q\rangle}$ reads
\begin{eqnarray}
 \Delta\Phi_V(M^2, Q^2)
  = \frac{8\,A_0}{M^4}\,
     \left(2+\frac{Q^2}{2\,M^2-\lambda_q^2}\right)
      \exp\left[\frac{-\,Q^2\,\lambda_q^2}
                     {2\,M^2\left(2\,M^2-\lambda_q^2\right)}
          \right]\,.
\end{eqnarray}
This term indeed vanishes for large $Q^2$ and is controlled
by the nonlocality parameter $\lambda_q^2$.
The larger $\lambda_q^2$,
the faster this contribution decreases with $Q^2$.
The explicit expressions for the other condensate contributions are
omitted here, but their schematic $Q^2$ dependence can be read off
from Table \ref{tab:History}.

\begin{figure}[t]
 \centerline{\includegraphics[width=0.47\textwidth]{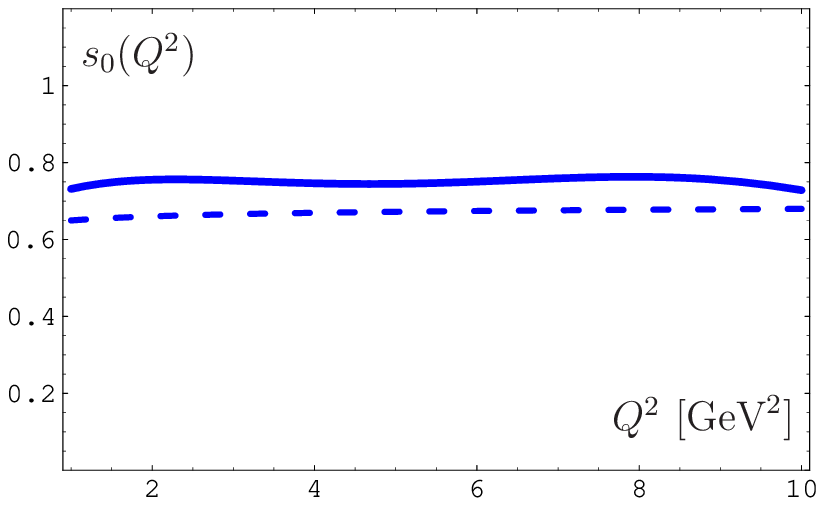}~~~
             \includegraphics[width=0.47\textwidth]{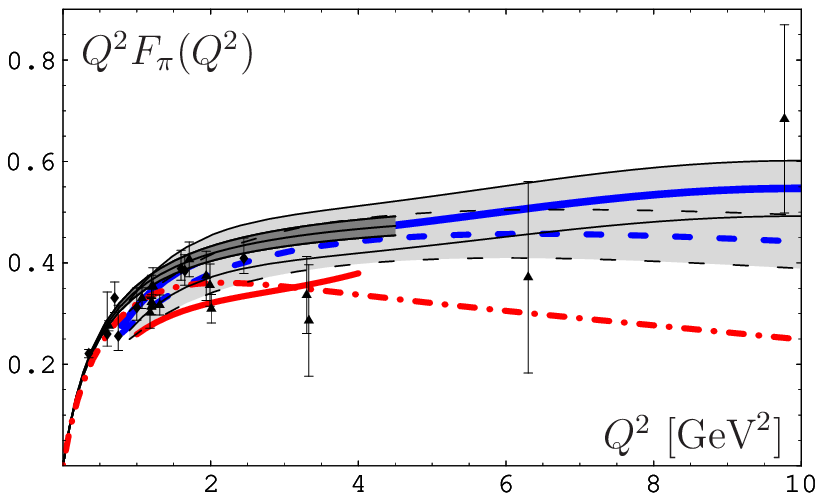}}
  \caption{\label{fig:1}\footnotesize
      Left panel:
      Continuum threshold $s_0(Q^2)~[\text{GeV}^2]$ for the minimal
      (dashed line) and for the improved (solid line) NLC model.
      Right panel:
      Theoretical predictions for scaled pion form factor
      $Q^2 F_{\pi}(Q^2)$ obtained by different methods and models.
      Dashed line---minimal NLC model;
      solid line---improved NLC model (in both cases
      $\lambda_q^2=0.4~\text{GeV}^2$ has been used and the corresponding
      uncertainties are indicated by light gray strips delimited by similar lines).
      The following designations are used:
      thick line between 1 and 4~GeV$^2$---standard QCD SR
      with local condensates~\cite{NR82,IS82};
      dash-dotted line---LD QCD SR~\cite{BLM07};
      triangles---Cornell experimental data \cite{FFPI78};
      diamonds---JLab experimental data \cite{JLab08II}.
      The recent lattice result \cite{Brommel06} is shown as a monopole
      fit containing error bars illustrated by a dark grey strip
      ending at $\approx4.5$~GeV$^2$.
      }
   \end{figure}

The pion form factor $F_{\pi}(M^2,s_0)$, as a function of two additional
parameters $M^2$ (Borel parameter) and $s_0$ (continuum threshold),
is given at each fixed value of $Q^2$ by SR Eq.\
(\ref{eq:ffQCDSR}).
The parameter $s_0$ marks the boundary between the pion state
and
higher resonances ($A_1$, $\pi'$, etc.).
We select its value at each momentum transfer $Q^2$
by applying to the function $F_{\pi}(M^2,s_0)$
the minimal-sensitivity condition with
respect to the auxiliary parameter $M^2$ in the fiducial interval of
the SR.
These intervals and the value of the pion decay constant $f_{\pi}$
for each considered NLC model, notably the minimal
and the improved Gaussian one, are taken from the corresponding 2-point
NLC QCD SR:
$f_{\pi}=0.137$~GeV$^2$,
$M^2_{-}=1$~GeV$^2$, and
$M^2_{+}=1.7$~GeV$^2$
for the minimal Gaussian model~\cite{BMS01},
whereas for the improved one~\cite{BP06}
one has
$f_{\pi}=0.142$~GeV$^2$,
$M^2_{-}=1$~GeV$^2$,
and
$M^2_{+}=1.9$~GeV$^2$.
The continuum threshold $s_0^{\text{SR}}(Q^2)$, which
minimizes the dependence of the right-hand side of (\ref{eq:ffQCDSR}),
is fixed by the root-mean-square deviation
$\chi^2(Q^2,s_0)$, Eq.\ (\ref{eq:xi}),
in the Borel-parameter interval
$M^2\in[M^2_{-}, M^2_{+}]$ at each value of $Q^2$.
The results of this procedure are shown in the left panel of
Fig.~\ref{fig:1}.
Both models generate approximately constant values
of $s_0(Q^2)$ in the whole region $Q^2\in[1;10]$~GeV$^2$
with slightly higher values in the case of the improved Gaussian model.

The SR result for the pion form factor
is defined numerically as
the average value of the right-hand side of SR (\ref{eq:ffQCDSR})
with respect to the Borel parameter $M^2\in[M^2_{-}, M^2_{+}]$:
\begin{eqnarray}
 \label{eq:FF.pi.SR.result}
 F_{\pi}^\text{SR}(Q^2)
  &=& \frac{1}{M_{+}^2-M_{-}^2}
       \int_{M_{-}^2}^{M_{+}^2}
        F(Q^2, M^2, s_0(Q^2))\,dM^2\,.
\end{eqnarray}
The obtained predictions for the pion form factor
for both Gaussian NLC models
with $\lambda_q^2=0.4$~GeV$^2$,
are shown in the right panel of Fig.~\ref{fig:1}
as dashed and solid curves,
respectively, in comparison with the lattice result
of \cite{Brommel06}
(dark grey strip limited from above at approximately $5$~GeV$^2$).
These theoretical results are compared with the available experimental
data~\cite{FFPI78,JLab08II} and previous theoretical estimates
\cite{NR82,IS82,BLM07}.
We also show in this figure in the form of light grey strips
the minimal theoretical uncertainties of the QCD SR results.
The central curves of our predictions can be represented by
the corresponding interpolation
formulas:
\begin{subequations}
\begin{eqnarray}
 \label{eq:FF.pi.SR.Int.Min}
  F_{\pi;\text{Min}}^{\text{SR}}(Q^2=x~\text{GeV}^2)
   &=& 1.64\,e^{-1.73\,x^{0.32}} x\,,\\
 \label{eq:FF.pi.SR.Int.Imp}
  F_{\pi;\text{Imp}}^{\text{SR}}(Q^2=x~\text{GeV}^2)
   &=& e^{-0.528\,x^{0.8}} x \left(0.016\,x^2-0.065\,x+0.58\right)\,,
\end{eqnarray}
\end{subequations}
valid for $Q^2\in[1,10]$~GeV$^2$, i.~e., for $x\in[1,10]$.

\section{Discussion and Conclusions}
\label{sec:Conclusions}
We calculated the electromagnetic pion FF
 using QCD SR with NLC~\cite{MR86,BR91}
 with a QCD vacuum nonlocality parameter $\lambda_q^2=0.4~\text{GeV}^2$
 and  using a perturbative spectral density proposed in \cite{BO04}.
This $\lambda_q^2$ value receives support from a recent comprehensive
analysis \cite{BMS02,Ste08} of the CLEO data on the pion-photon
transition.\footnote{Using somewhat higher values of this parameter,
would entail a decrease of the pion form factor owing to a stronger
influence of the nonlocality effects.
The opposite effect appears for smaller values of this parameter and
leads to an increase of the pion form factor.}
The use of NLC enables one to considerably enlarge the region of
applicability of the QCD SR towards momenta as high as
$10~\text{GeV}^2$---in contrast to the standard QCD SR approach \cite{NR82,IS82},
where the SR can be applied only in the $Q^2\leq3$~GeV$^2$ region.

The main conclusions of our investigation can be summarized as follows:
\begin{itemize}
  \item We found that the $O(\alpha_s)$-contribution influences
   the pion form factor at the level of $20\%$.
   This estimate is a little bit smaller than ones,
   obtained in~\cite{BO04,BLM07}.
  \item We found that the central-line prediction of the improved model NLC
   model is inside the error strip of the minimal model up to $Q^2=6$~GeV$^2$.
   Therefore, we may conclude that both models are equally good
   in this region.
   In view of the absence of more precise experimental data
   on the pion form factor at present,
   we cannot give any preference to one or the other
   of the two considered NLC models.
  \item It appears that our predictions are systematically higher
   than those of the LD approach~\cite{BLM07}.
   This can be easily understood in terms of the effective LD threshold
   $s_0^\text{LD}(Q^2)$.
   As we have already said, its value in the LD approach is well
   established only in the small-$Q^2$ region.
   For higher values, it is not firmly fixed; for instance, in
   \cite{BLM07} it was suggested to use a logarithmically increasing
   threshold
   \begin{eqnarray*}
    s_0^\text{LD}(Q^2)
   =
    \frac{4\pi^2 f_\pi^2}{1+\alpha_s(Q^2)/\pi}\,,
   \end{eqnarray*}
   which is $\approx0.62$~GeV$^2$ for $Q^2\approx3$~GeV$^2$.
   We estimated that in order to imitate the NLC QCD SR results
   in the LD approach, one needs to use
   $s_0^\text{LD}(Q^2=10~\text{GeV}^2)=0.87~\text{GeV}^2$.
   This means that the $s_0^\text{LD}$ uncertainty in the region of
   $Q^2=10$~GeV$^2$ is of the order of $30$\%
   This is the origin of the discussed difference between the LD
   results and ours.
  \item The lattice QCD results of~\cite{Brommel06} are in excellent
   agreement with our predictions.
  \item Both, the minimal and the improved Gaussian model for the NLC
   give results which are in good agreement within errors with the
   currently available experimental data.
\end{itemize}

\section*{Acknowledgements}
We are grateful to S.~V.~Mikhailov for helpful discussions.
Two of us (A.~P.~B. and A.~V.~P.) are indebted to Prof.\ Klaus Goeke
for the warm hospitality at Bochum University, where part of this
work was done.
This work was supported in part by the Heisenberg--Landau Programme,
grant 2009, the Russian Foundation for Fundamental Research,
grants No.\ ü~07-02-91557, 08-01-00686, and 09-02-01149,
and the BRFBR--JINR Cooperation Programme, contract No.\ F08D-001.

\begin{appendix}

\section{QCD SR parameters}
 \renewcommand{\theequation}{\thesection.\arabic{equation}}
  \label{App:A.0}\setcounter{equation}{0}
The parameters of the NLC are
$\Lambda=\lambda_q^2/2=0.2$~GeV$^2$,
$\langle\alpha_s{GG}\rangle/\pi=0.012$~GeV$^4$,
and
$\alpha_s\,\langle\bar{q}q\rangle^2$ = $1.83\cdot 10^{-4}$~GeV$^6$.
The nonlocal gluon-condensate contribution
$\Phi_{G}(M^2)$
produces a very complicated expression.
In analogy to the quark case, we model it by an exponential factor
\cite{BR91,MS93}:
$\Phi_{G}(M^2)=\Phi_{G}^\text{loc}(M^2)\,e^{-\lambda_g^2 Q^2/M^4}$
with $\lambda_g^2=0.4$~GeV$^2$.

In order to determine the best value of the threshold $s_0$,
we define the $\chi^2$ function for each value of $Q^2$ and $s_0$
as follows
\begin{eqnarray}\label{eq:xi}
 \chi^2(Q^2,s_0)
  = \frac{\varepsilon^{-2}}{N_M}
    \left[\sum\limits_{i=0}^{N_M}
           Q^4\,F(Q^2,M^2_{i},s_0)^2
        - \frac{\left(\sum\limits_{i=0}^{N_M}
                       Q^2\,F(Q^2,M^2_{i},s_0)
                \right)^2}
               {N_M+1}
    \right]\,,
\end{eqnarray}
where we used $M_{i}^2=M_{-}^2+i\,\Delta_M$,
$\Delta_M=(M_{+}^2-M_{-}^2)/N_M$,
$N_M=10$,
and with $\varepsilon$ denoting the desired accuracy for
$\chi^2\simeq1$
(the actual value used in the computation is $\varepsilon=0.07$.)
\end{appendix}


\end{document}